\documentclass{ws-ijmpe}

\begin{document}

\markboth{R.B. NEUFELD}{TAGGED JETS AND JET RECONSTRUCTION AS A PROBE OF QGP INDUCED PARTONIC ENERGY LOSS}

\catchline{}{}{}{}{}
%%%%%%%%%%%%%%%%%%%%%%%%%%%%%%%%%%%%%%%%%%%%%%%%%%%%%%%%%%%%%%%%%%%%

\title{TAGGED JETS AND JET RECONSTRUCTION AS A PROBE OF QGP INDUCED PARTONIC ENERGY LOSS
}

\author{\footnotesize R.B. NEUFELD}

\address{Theoretical Division, Los Alamos National Lab\\
Los Alamos, New Mexico, USA,
Country\\
neufeld@lanl.gov}

\maketitle

\begin{history}
\received{(received date)}
\revised{(revised date)}
%\accepted{(Day Month Year)}
%\comby{(xxxxxxxxxx)}
\end{history}

\begin{abstract}
Recent experimental advances at the Relativistic Heavy Ion Collider (RHIC) and the large center-of-mass energies available to the heavy-ion program at the Large Hadron Collider (LHC) will enable strongly interacting matter at high temperatures and densities, that is, the quark-gluon plasma (QGP), to be probed in unprecedented ways. Among these exciting new probes are fully-reconstructed inclusive jets and the away-side hadron showers associated with a weakly or electromagnetically interacting boson, or, tagged jets. Full jet reconstruction provides an experimental window into the mechanisms of quark and gluon dynamics in the QGP which is not accessible via leading particles and leading particle correlations. Theoretical advances in this growing field can help resolve some of the most controversial points in heavy ion physics today. I here discuss the power of jets to reveal the spectrum of induced radiation, thereby shedding light on the applicability of the commonly used energy loss formalisms and present results on the production and subsequent suppression of high energy jets tagged with Z bosons in relativistic heavy-ion collisions at RHIC and LHC energies using the Gyulassy-Levai-Vitev (GLV) parton energy loss approach.
\end{abstract}

$$$$

The suppression of energetic partons in the QGP, or {\it jet quenching}, is one of the most important results from the heavy-ion program at RHIC \cite{Gyulassy:2003mc,Jacobs:2004qv}.  The limited center-of-mass energies available at RHIC have mostly restricted experimental measurements thus far to leading particle suppression relative to $p+p$ collisions \cite{Adler:2006hu}.  Unfortunately, leading particle suppression alone cannot discriminate between partonic energy loss formalisms or extract quantitatively the jet quenching properties of the QGP \cite{Bass:2008rv}.

\begin{figure}[th]
\centerline{\psfig{file=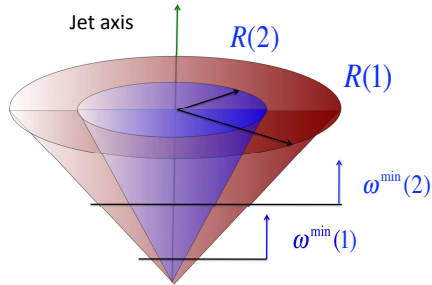,width=5cm,angle=270}}
\vspace*{8pt}
\caption{A schematic illustration of the jet reconstruction cone radius, $R = \sqrt{(\Delta \phi)^2 + (\Delta y)^2}$, and the particle tower energy selection, $\omega_{min}$.  As these parameters are varied, the amount of energy recovered in jet reconstruction (and hence the jet cross section) also varies.}
\label{one}
\end{figure}

In order to constrain the underlying quantum chromodynamic (QCD) theory of jet quenching, new and more differential observables are needed.  Particularly promising are jet shapes \cite{Vitev:2008rz,Vitev:2009rd} and jets tagged with electromagnetic \cite{Neufeld:2010fj,Srivastava:2002kg} or weakly interacting probes. Jet reconstruction and analysis of the subsequent jet shapes provide an experimental handle on the large angle medium-induced bremsstrahlung spectrum \cite{Vitev:2005yg}.  To understand why this is true, consider Fig. \ref{one}
\begin{figure}[th]
\centerline{\psfig{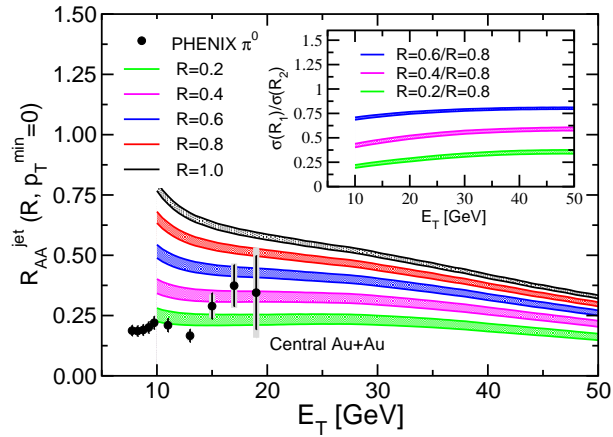}}
\vspace*{8pt}
\caption{The figure gives predictions for jet $R_{AA}$ as a function of jet reconstruction cone size $R = \sqrt{(\Delta \phi)^2 + (\Delta y)^2}$ using the GLV partonic energy loss formalism.  The plot is for impact parameter b = 3 fm in Au+Au collisions at $\sqrt{s} = 200$ GeV per nucleon.  The sensitivity of the suppression on cone size $R$ is reflective of the angular distribution of medium-induced radiation.  The insert shows the ratios of jet cross sections for selected values of $R$.}
\label{two}
\end{figure}
from Ref. \cite{Vitev:2008rz}, which shows a schematic illustration of the jet reconstruction cone radius, $R = \sqrt{(\Delta \phi)^2 + (\Delta y)^2}$, and the particle tower energy selection, $\omega_{min}$.  As one varies these parameters, the amount of energy recovered in jet reconstruction (and hence the jet cross section) likewise varies.  This variation can be exploited to reveal the structure of the underlying medium induced bremsstrahlung spectrum, as is demonstrated in Fig. \ref{two}, which is taken from Ref. \cite{Vitev:2009rd}.  The Figure shows predictions for the sensitivity of jet $R_{AA}$ on jet reconstruction cone size $R$ using the Guylassy-Levai-Vitev \cite{Gyulassy:2000fs} (GLV) partonic energy loss formalism.  The plot was done for impact parameter b = 3 fm in Au+Au collisions at $\sqrt{s} = 200$ GeV per nucleon.  The dependence of the suppression on cone size $R$ directly reflects the angular distribution of medium-induced radiation.  The heavy-ion program at the LHC and the future RHIC
upgrades will enable a high statistics experimental measurement of this feature for the first time.

{\it Tagged jets} also provide an exciting opportunity to quantify the jet quenching properties of the QGP.  Tagged jets are high energy partons produced in association with the tagging particle.  Electroweak bosons are ideal tags to study partonic energy loss because, once produced, they have negligible medium induced modifications.  In the limit of leading order (LO) kinematics, the tagging particle serves as a signal for the initial associated jet energy.  In this way, one hopes to obtain the amount and distribution of partonic energy loss by reconstructing the tagged jet.  The leptonic final states of the $Z^0$ boson are an especially attractive jet tag because the large invariant mass of the $Z^0$ boson makes it easy to distinguish from the background generated in a heavy-ion collision.  The heavy-ion program at the LHC will enable the experimental measurement of this feature for the first time.

As mentioned above, at LO the kinematics ensure that the tagging $Z^0$ boson serves as an exact measure of the associated jet $p_T$ (magnitude of momentum transverse to the beam line).
\begin{figure}[th]
\vskip0.04\linewidth
\centerline{\psfig{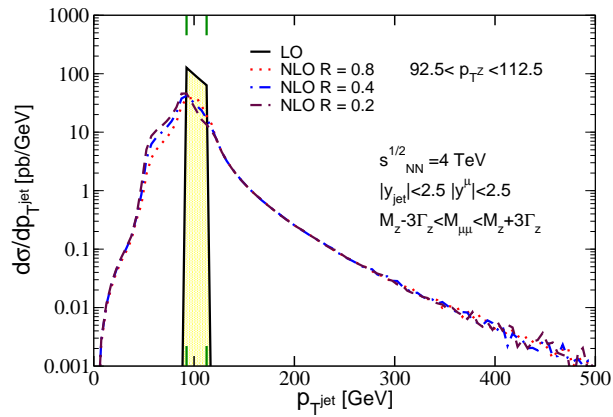}}
\vspace*{2pt}
\caption{When going from LO to NLO, the tagging power of the $Z^0$ boson is compromised by the possibility of an extra jet in the final state, as seen from the figure.  The LO result is constrained exactly by the $p_T$ cut on the tagging particle.  However, this is no longer true at NLO.}
\label{three}
\end{figure}
This is demonstrated in Fig. \ref{three} where I present results \cite{Neufeld:2010fj} for jets associated with $Z^0$/$\gamma^*\rightarrow \mu^+\mu^-$ in p+p collisions at $\sqrt{s} = 4$ TeV.  In the plot, the tagging $Z^0$/$\gamma^*$ is required to have $92.5 < p_T < 112.5$ GeV and curves are presented for the leading and next-to-leading order (NLO) results, as well as for three different values of jet reconstruction radius.  The NLO calculations are performed using the publicly available Monte Carlo for FeMtobarn processes (MCFM) developed by Campbell and Ellis \cite{Campbell:2002tg}.  In the figure, one indeed sees that at LO the $Z^0$/$\gamma^*$ serves as an exact tag of the associated jet energy, as the jet $p_T$ distribution is entirely inside of the tagging window $92.5 - 112.5$.  However, at NLO this correspondence is largely washed away due to the possibility of an extra jet in the final state.  This relatively large smearing effect at NLO suggests great care must be taken in experimentally tagging the initial associated jet energy in $AA$ collisions.

\begin{figure}[th]
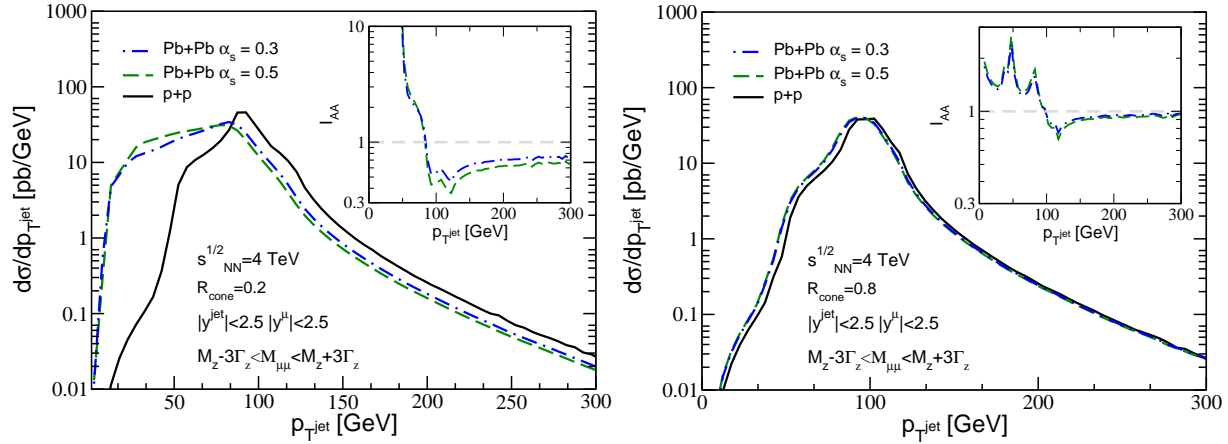

\centerline{\psfig{file=925_1125_r02.eps,width=8cm}
\psfig{file=925_1125_r08.eps,width=8cm}}
\vspace*{8pt}
\caption{The NLO $p_T$ differential cross section per nucleon pair for jets tagged with $Z$/$\gamma^*\rightarrow \mu^+\mu^-$ in $p+p$ and $Pb+Pb$.  The tagging $Z$/$\gamma^*$ is required to have $92.5 < p_T < 112.5$.  The results are shown for two different values of jet cone radius $R = 0.2$ (left panel) $R = 0.8$ (right panel).  The medium induced energy loss tends shifts the curve to lower values of $p_T$.  As the jet cone radius is increased, the medium modified curves approach the $p+p$ result, as more and more of the medium induced bremsstrahlung is recovered in the jet.}
\label{four}
\end{figure}

I next consider results for jets associated with $Z^0$/$\gamma^*\rightarrow \mu^+\mu^-$ in Pb+Pb collisions at $\sqrt{s} = 4$ TeV/nucleon pair.  In this proceedings I will not consider cold nuclear matter effects \cite{Vitev:2007ve}, but will leave that for a future study.  In Pb+Pb collisions it is essential to include jet energy loss through medium-induced bremsstrahlung.  The partonic energy loss in medium for the results that follow was obtained from the GLV energy loss formalism \cite{Gyulassy:2000fs}.  The NLO results for tagging $Z^0$/$\gamma^*$ required to have $92.5 < p_T < 112.5$ GeV is shown in Fig. \ref{four} \cite{Neufeld:2010fj}.
The result is shown for two different values of jet cone reconstruction radius, $R = 0.2, 0.8$ (left and right panel, respectively).  The $p_T$ differential cross section per nucleon pair for Pb+Pb is superimposed with the result from p+p which was shown in Fig. \ref{three}.  As is clear from the figure, the effect of medium induced energy loss is to shift the curve to lower values of $p_T$.  It is also clear from the figure that as the jet cone radius is increased, the medium modified curves approach the $p+p$ result, as more and more of the medium induced bremsstrahlung is recovered in the jet.  This occurs regardless of the strength of the medium coupling, $\alpha_s$.  As the jet cone radius is decreased, the sensitivity to $\alpha_s$ becomes more important, as less of the medium induced radiation is recovered.  This sensitivity of the result in Pb+Pb to jet cone radius is directly related to the angular distribution of the medium-induced bremsstrahlung spectrum.

In summary, recent experimental advances at RHIC and the upcoming heavy-ion program at the LHC will usher in a new era of experimental capabilities to probe the quark-gluon plasma.  Among these new probes, jet shapes and jets tagged with weakly interacting probes are especially promising, particularly in light of the fact that leading particle suppression alone is not sufficient to discriminate between partonic energy loss formalisms or to extract quantitatively the jet quenching properties of the QGP.

I have here discussed the power of jets to reveal the spectrum of induced radiation, thereby shedding light on the applicability of the commonly used energy loss formalisms and presented results on the production and subsequent suppression of high energy jets tagged with Z bosons in relativistic heavy-ion collisions using the GLV parton energy loss approach.  The results suggest that higher order production processes will make experimentally tagging the initial associated jet energy with electroweak bosons difficult.  It is also observed that the dependence of the suppression in Pb+Pb relative to p+p on jet reconstruction radius $R$ is a sensitive probe of the spectrum of medium induced radiation.  It is worthwhile to here mention that tagged jets may provide an exciting opportunity to probe the medium response to hard partons \cite{Neufeld:2009ep}, especially if the NLO smearing effects are brought under sufficient theoretical control.

\section*{Acknowledgements}

I wish to thank my collaborators Ivan Vitev and Ben-Wei Zhang, and also Jaroslav Bielcik and Jana Bielcikova for hosting an excellent workshop.  This work was supported in part by the US Department of Energy, Office of Science, under Contract No. DE-AC52-06NA25396.

\appendix

\end{document}